\documentclass{aa}
\usepackage{natbib}
\usepackage{txfonts}
\usepackage{graphicx}
\usepackage{subfigure}
\usepackage{exscale}
\usepackage[english]{babel}
\bibpunct{(}{)}{;}{a}{}{,}

\begin{document}

\title{\object{DE CVn}: A bright, eclipsing red dwarf - white dwarf
binary}

\author{E.J.M.~van~den~Besselaar\inst{1} \and R.~Greimel\inst{2} \and
 L.~Morales-Rueda\inst{1} \and G.~Nelemans\inst{1} \and
 J.R.~Thorstensen\inst{3} \and T.R.~Marsh\inst{4} \and
 V.S.~Dhillon\inst{5} \and R.M.~Robb\inst{6} \and D.D.~Balam\inst{6}
 \and E.W.~Guenther\inst{7} \and J.~Kemp\inst{8} \and
 T.~Augusteijn\inst{9} \and P.J.~Groot\inst{1} }

\offprints{E.J.M. van den Besselaar, \\ \email{besselaar@astro.ru.nl}}

\institute{
Department of Astrophysics, IMAPP, Radboud University Nijmegen, PO
Box 9010, 6500 GL Nijmegen, The Netherlands; \\
\email{[besselaar;lmr;nelemans;pgroot]@astro.ru.nl} 
\and Isaac Newton Group of
Telescopes, Apartado de correos 321, E-38700 Santa Cruz de la
Palma, Spain; \email{greimel@ing.iac.es} 
\and Department of Physics and Astronomy, Dartmouth College, 6127
Wilder Laboratory Hanover, NH 03755, USA;
\\\email{thorsten@partita.dartmouth.edu}
\and Department of Physics, University of Warwick, Coventry CV4 7AL,
UK; \email{t.r.marsh@warwick.ac.uk}
\and Department of Physics and Astronomy, University of Sheffield,
Sheffield S3 7RH, UK; \email{vik.dhillon@sheffield.ac.uk}
\and Department of Physics and Astronomy, University of Victoria,
Victoria, BC, V8W 3P6, Canada; \\ \email{robb@uvic.ca; cosmos@uvvm.uvic.ca}
\and Th\"uringer Landessternwarte Tautenburg, Sternwarte 5, D-07778
Tautenburg, Germany; \email{guenther@tls-tautenburg.de}
\and Joint Astronomy Centre 660 N. A'ohoku Place University Park Hilo,
Hawaii 96720, USA; \email{j.kemp@jach.hawaii.edu}
\and Nordic Optical Telescope, Apartado 474, E-38700 Santa Cruz de La
Palma, Spain; \email{tau@not.iac.es} }

\date{Received 15 August 2006 / Accepted 16 January 2007}

\abstract{Close white dwarf - red dwarf binaries must have gone
through a common-envelope phase during their
evolution. \object{DE~CVn} is a detached white dwarf - red dwarf
binary with a relatively short ($\sim$$8.7$ hours) orbital period. Its
brightness and the presence of eclipses makes this system ideal for a
more detailed study.}
%
{From a study of photometric and spectroscopic observations of
\object{DE~CVn} we derive the system parameters which we discuss in
the frame work of common-envelope evolution.}
%
{Photometric observations of the eclipses are used to determine an
accurate ephemeris. From a model fit to an average low-resolution
spectrum of \object{DE~CVn} we constrain the temperature of the white
dwarf and the spectral type of the red dwarf. The eclipse light curve
is analysed and combined with the radial velocity curve of the red
dwarf determined from time-resolved spectroscopy to derive constraints
on the inclination and the masses of the components in the system.}
%
{The derived ephemeris is
HJD$_\mathrm{min}$~=~2452784.5533(1)~+~0.3641394(2)~$\times$~E.  The
red dwarf in \object{DE~CVn} has a spectral type of M3V and the white
dwarf has an effective temperature of
$8\,000$~K. The inclination of the system is $86^{+3 \circ}_{-2}$ and the
mass and radius of the red dwarf are $0.41\pm0.06$~$M_{\odot}$
and $0.37^{+0.06}_{-0.007}$~$R_{\odot}$, respectively, and the
mass and radius of the white dwarf are
$0.51^{+0.06}_{-0.02}$~$M_{\odot}$ and
$0.0136^{+0.0008}_{-0.0002}$~$R_{\odot}$, respectively.}
{We found that the white dwarf has a hydrogen-rich atmosphere
(DA-type). Given that \object{DE~CVn} has experienced a
common-envelope phase, we can reconstruct its evolution and we find
that the progenitor of the white dwarf was a relatively low-mass star
(M$\leq1.6M_{\odot}$). The current age of this system is
$3.3-7.3\times10^9$ years, while it will take longer than the Hubble
time for \object{DE~CVn} to evolve into a semi-detached system.}


\maketitle

\section{Introduction}
\label{sec:introduction}
Large gaps remain in our knowledge of binary stellar evolution that
affect our understanding of not only evolved compact binaries, but
also of phenomena such as supernovae type Ia explosions, the rate of
neutron star -- neutron star mergers, and the number of gravitational
wave sources in our Galaxy. The poorly understood physics of the
common-envelope (CE) phase results in considerable uncertainty on the
binary evolution \citep{paczynski}. During the evolution of a binary,
the more massive star turns into a giant. When the initial orbital
period is small enough \citep[$\lesssim$10 years,][]{taam} the
envelope of the giant will encompass the secondary star. The secondary
and the core of the giant will spiral in towards each other in a
common-envelope. When the envelope is expelled a close binary
consisting of the core of the giant, which will evolve towards a white
dwarf, and the un-evolved secondary star may emerge
\citep[see e.g.][]{nelemans}.

The common-envelope phase is expected to be very short
\citep[$\lesssim 1000$ years,][]{taam} and is therefore virtually
impossible to observe directly. To study the effects of this phase it
is best to focus on objects that have most probably undergone a
common-envelope phase in their past. These we identify with binary
systems containing at least one stellar remnant where the current
orbital separation is smaller than the radius of the giant progenitor
(usually with orbital periods $\leq 1$ day).

Eclipsing close binaries offer the greatest possibility of deriving
precise physical parameters of the stars. The masses, radii and
orbital separations give insight into the binary evolution and
specifically tell us if a CE phase has happened sometime in their
past.

Some examples of detached, close white dwarf - low-mass main-sequence
star (red dwarf) eclipsing binaries are: \object{RR~Cae}
\citep{rrcae}, \object{NN~Ser} \citep{nnser}, \object{EC13471-1258}
\citep{ec13471}, \object{GK~Vir} \citep{gkvir} and \object{RX
J2130.6+4710} \citep{rxj2130}. For a review on detached white dwarf -
red dwarf binaries see e.g. \citet{wdrdreview} and
\citet{schreiber}. The latest list of white dwarf - red dwarf binaries
is in \citet[][with ten new systems compared to
\citealt{wdrdreview}]{morales}. It is necessary to study as many of
these systems as possible to be able to compare their characteristics
with population synthesis models and to find their space densities as
a function of composition (e.g. white dwarf temperature, spectral
type, age).

\object{DE~CVn} (\object{RX~J1326.9+4532}) is a relatively unstudied,
bright ($V=12.8$) eclipsing binary. It was first discovered as an
X-ray source by ROSAT \citep{rosat} and has a proper motion of $-0.198
\pm 0.002 \arcsec$~yr$^{-1}$ in right ascension and $-0.178 \pm 0.003
\arcsec$~yr$^{-1}$ in declination as given in the USNO-B1 catalog
\citep{usnob}.

This object was first studied photometrically by \citet{robert}. From
the light curve and the unequal maxima they derived an orbital period
of $0.364$ days. The asymmetry in their light curve needed a star spot
to accurately model the light curve. \citet{robert} measured eclipse
depths of $0.054\pm0.010$ magnitude in the $R$ band and
$0.128\pm0.029$ magnitude in the $V$ band.

\citet{holmes} obtained $UBVRI$ photometry for five nights in June
2000. They confirm the dependence of the eclipse depth with colours
and found minimum depths of the eclipse of $0.10$ magnitude in $I$,
$0.15$ in $R$, $0.30$ in $V$, $0.60$ in $B$, and $1.00$ in $U$. The
differences with wavelength band indicate that the two stars have very
different colours.

We note a difference in the eclipse depths as quoted by \citet{holmes}
compared to the values from \citet{robert}. Although \citet{holmes}
give their values as being eclipse depths, when looking at the light
curve we suggest that they have taken the difference between minimum
and maximum light instead of the difference between the start and
minimum of the eclipse which is used by \citet{robert} and in the
present work.

\object{DE~CVn} consists of an M-type star with a spectroscopically
unseen companion, presumably a white dwarf. Throughout this paper we
will refer to the M dwarf as the secondary component and the probable
white dwarf as the primary component.

In Sect.~\ref{sec:observations} we describe our observations and
reductions. The results are shown in Sect.~\ref{sec:results} and the
conclusions are given in Sect.~\ref{sec:conclusions}.

\section{Observations and reductions}
\label{sec:observations}
\subsection{Photometry}
\label{sec:obs:photo}
Our photometric dataset consists of various observations taken on a
number of telescopes. Table~\ref{tab:photolog} lists an overview of
our photometric datasets taken with the 1.3-meter telescope of the
Michigan-Dartmouth-MIT (MDM) Observatory (Arizona), with the 4.2-meter
William Herschel Telescope (WHT) on La Palma with ULTRACAM
\citep{ultracam}, with the automatic 0.5-meter telescope of the
Climenhage Observatory in Victoria, Canada (referred to as UVic) and
with the 1.8-meter telescope of the Dominion Astrophysical Observatory
(DAO) located in Victoria, Canada.

\begin{table}[!t]
  \caption{Log of the photometric data of \object{DE~CVn}.  UVic is
  the automatic 0.5-meter telescope of the Climenhage Observatory in
  Victoria, Canada. DAO is the 1.8-meter telescope of the Dominion
  Astrophysical Observatory. MDM is the 1.3-meter telescope of the
  Michigan-Dartmouth-MIT Observatory in Arizona. ULTRACAM are
  observations with this instrument at the WHT. T is the integration
  time per observation in seconds and \# is the number of
  observations.}
  \label{tab:photolog}
  \begin{tabular}{r l l r r}
\hline\hline
    Date & Tel & Filter & T & \# \\
    \hline
12 Apr 1997  & UVic     & $R$       &  99   & 173 \\
21 Apr 1997  & UVic     & $R$       &  99   & 151 \\
22 Apr 1997  & UVic     & $R$       & 120   & 179 \\
24 Apr 1997  & UVic     & $R$       & 140   & 125 \\
 2 May 1997  & UVic     & $R$       &  99   & 175 \\
 8 May 1997  & UVic     & $R$       &  99   &  76 \\
 8 May 1997  & UVic     & $V$       & 120   &  76 \\
 9 May 1997  & UVic     & clear     &  34   & 164  \\
10 May 1997  & UVic     & clear     &  33   & 484  \\
 1 Jul 1998  & UVic     & $R$       & 140   & 101  \\
 7 Mar 2000  & UVic     & clear     &  99   &  52  \\
21 May 2001  & UVic     & $R$       & 120   & 169  \\
26 May 2001  & DAO      & $B$       &  40   & 110  \\
 2 Jun 2001  & UVic     & clear     &  40   & 130  \\
21 Jan 2002  & MDM      & $B$       &  60   & 193  \\
24 Jan 2002  & MDM      & $BG38$    &   4   & 597  \\
30 May 2002  & DAO      & $B$       &  30   &  39  \\
 1 Jul 2002  & MDM      & $BG38$    &   5   & 216  \\
 3 Feb 2003  & MDM      & $B$       &  10   & 350  \\
22 May 2003  & ULTRACAM & $u'g'i'$  & 1.3 & $\sim$750 \\
24 May 2003  & ULTRACAM & $u'g'i'$  & 1.3 & $\sim$2860 \\
25 May 2003  & ULTRACAM & $u'g'i'$  & 1.3 & $\sim$3780 \\
19 Jun 2003  & MDM      & $B$       & 10  & 166 \\
 4 May 2004  & ULTRACAM & $u'g'i'$  & 1.3 & $\sim$13000 \\
 5 Apr 2006  & UVic     & clear     & 33  & 359 \\
\hline
 \end{tabular}
\end{table}

To reduce dead-time for the MDM observations, the CCD was binned
$2\times2$, resulting in a scale of $1.02\arcsec$~per binned pixel, and
only a subregion of $256\times256$ (binned) pixels was read. These
observations were reduced with standard packages in the Image
Reduction and Analysis Facility (IRAF)\footnote{IRAF is distributed by
the National Optical Astronomy Observatories, which are operated by
the Association of Universities for Research in Astronomy, Inc., under
cooperative agreement with the National Science Foundation.}. We
derived differential photometry with respect to one comparison star
(RA = 13:26:59.6, Dec = +45:33:05, J2000) that is bright enough in the
sparse field. The UVic and DAO observations were reduced with standard
packages in IRAF. 

The ULTRACAM data were reduced with standard aperture
photometry. Differential photometry was obtained with respect to two
comparison stars located at RA = 13:26:28.08, Dec = +45:33:11.6
(J2000) and RA = 13:26:39.2, Dec = +45:34:56.1 (J2000). The
coordinates of \object{DE~CVn} are RA = 13:26:53.2, Dec = +45:32:46.1
(J2000).

\begin{table}[!t]
  \caption{Log of spectroscopic observations where we have given the
  date, telescope (Tel), the wavelength range ($\lambda$) in \AA, the
  integration time (T) in s and the resolution (R) in \AA~as derived
  from the FWHM of the arc lines. TLS = 2m telescope of the
  Th\"uringer Landessternwarte 'Karl Schwarzschild' Tautenburg Echelle
  Spectrograph, Modspec = 2.4m Hiltner Telescope (MDM) Modular
  Spectrograph, CCDS = 2.4m Hiltner Telescope CCD Spectrograph, MkIII
  = 2.4m Hiltner Telescope Mark III Spectrograph, DAO = Cassegrain
  Spectrograph at the 1.8m of the DAO. }
  \label{tab:speclog}
  \begin{tabular}{r l l r r l}
\hline\hline
    Date         & Tel     & $\lambda$      &  T &  \# & R  \\
    \hline
11 May 1998      & TLS     & 4300--5300    & 900 & 21 & 0.14\\
12 - 18 May 1998 & TLS     & 5650--10\,000 & 900 & 75 & 0.16 \\
19 - 22 Jan 2002 & Modspec & 4250--7550    & 120 & 62 & 3.5\\
16 Feb 2002      & Modspec & 4250--7550    & 120 & 5  & 3.5\\
19 Feb 2002      & Modspec & 4250--7550    & 120 & 10 & 3.5\\
5 - 7 May 2002   & CCDS    & 4180--5100    & 300 & 18 & 3.2\\
13 May 2002      & CCDS    & 4180--5100    & 360 & 1  & 3.2\\
13 May 2002      & CCDS    & 4180--5100    & 600 & 1  & 3.2 \\
12 Jun 2002      & MkIII   & 4250--7550    & 90  & 1  & 3.7 \\
13 Jan 2004      & Modspec & 4250--7550    & 180 & 1  & 3.5\\
17 - 18 Jan 2004 & Modspec & 4250--7550    & 180 & 2  & 3.5 \\
6 Mar 2004       & Modspec & 4250--7550    & 180 & 2  & 3.5\\
30 Jan 2006      & ISIS    & 3000--8000    & 180 & 1  & 5\\
30 Jan 2006      & ISIS    & 3000--8000    & 240 & 6  & 5\\
15 May 2006      & DAO     & 3500--5150    & 720 & 6  & 4.8\\
\hline
  \end{tabular}
\end{table}

\subsection{Spectroscopy}
\label{sec:obs:spec}
A log of our spectroscopic data set is given in
Table~\ref{tab:speclog}. The TLS echelle spectra are reduced with
standard packages in IRAF. We could not correct the TLS spectra for
flat fielding, because the sky/twilight flat fields would introduce
only more lines in the spectra. The wavelength calibration was done
with $\sim$$300$ lines in the \ion{Th}{}-\ion{Ar}{} arc spectra with a
root-mean-square residual of $\sim$$0.0036$~\AA. We checked the
stability of the spectrograph by using sky lines and these are not
shifted between the observations. Because of the unavailability of sky
background or flux standards the echelle spectra are not sky
subtracted or flux-calibrated. These spectra were taken during grey
time, so we do not expect problems with the background level.

The MDM (using the Modspec, CCDS, MkIII spectrographs), DAO and ISIS
spectra were all reduced with standard packages in IRAF and these are,
except for the DAO spectra, approximately flux calibrated (a seeing
matched slit width was used, limiting the accuracy of the photometric
calibration).

\section{Results}
\label{sec:results}
\subsection{Eclipse light curves}
\label{sec:result:lightcurve}
The first ephemeris of the primary eclipse in \object{DE~CVn} is given
in \citet{robert}. We take the mid-eclipse times as the mid-point
between the start of the ingress and the end of the egress. From our
photometric observations and the times of minima given in
\citet{eclipsemin} we have determined the ephemeris of the eclipse
minima by fitting a straight line to the cycle numbers as derived
from the ephemeris of \citet{robert}:
\begin{equation}
 \mathrm{HJD}_{\mathrm{min}}= 2452784.5533(1) +
 0.3641394(2)\times\mathrm{E}
\end{equation}
No significant aliases were found near this period. The
uncertainties (last digits) are derived for $\Delta\chi^2=1$ when we
scale the individual errors to obtain that the reduced $\chi^2=1$. The
value of the orbital period is further confirmed by the radial
velocity analysis described in Sect.~\ref{sec:result:radvel}. The
availability of two eclipses separated by only two cycles ($-$6109 and
$-$6107) leaves no cycle count ambiguity. Two mid-eclipse times of Tas
et al. (2452411.3156 and 2452412.4078) were rejected due to the large
phase shifts ($\Delta\phi>0.01$, in contrast to an average scatter of
$0.004$ for all other eclipses) with respect to the updated ephemeris
of \citet{robert} leaving 23 mid-eclipse times for determining the
ephemeris. When calculating the phase difference with our new period,
it turns out that these two data points have indeed a large phase
difference ($\Delta\phi>0.01$), so most probably the published times
of minima are incorrect. The times of minima together with the cycle
number and the corresponding time difference are given in
Table~\ref{tab:mideclipse}.

\begin{table}[!t]
  \begin{center}
  \caption{Times of mid-eclipse for the binary \object{DE~CVn} are
  given together with the cycle number and the corresponding time
  difference. The numbers in parentheses are the uncertainties on the
  mid-eclipse times in the last digits.  $^a$~\citet{robert};
  $^b$~UVic photometry; $^c$~DAO photometry; $^d$~MDM photometry;
  $^e$~\citet{eclipsemin}; $^f$~ULTRACAM photometry; $^g$~Not used for
  determining the ephemeris.}
  \label{tab:mideclipse}
  \begin{tabular}{r@{.}l l r@{.}l }
\hline\hline
    \multicolumn{2}{c}{HJD-2450000}&cycle&\multicolumn{2}{c}{$\Delta$ (HJD)}\\ 
    \hline
550&9214(16)$^a$     & $-$6134 &  $-$0&00060 \\ 
560&0243(16)$^a$     & $-$6109 &  $-$0&00118 \\ 
560&7531(20$^a$)     & $-$6107 &  $-$0&00066 \\ 
562&9374(22)$^a$     & $-$6101 &  $-$0&00120 \\ 
570&9497(14)$^a$     & $-$6079 &     0&00004 \\ 
576&7749(14)$^a$     & $-$6063 &  $-$0&00100 \\ 
578&9606(6)$^a$      & $-$6057 &  $-$0&00013 \\ 
995&9003(18)$^b$     & $-$4912 &  $-$0&00010 \\ 
1620&7628(15)$^b$    & $-$3196 &  $-$0&00089 \\ 
2050&8125(16)$^b$    & $-$2015 &     0&00012 \\ 
2055&9103(7)$^c$     & $-$2001 &  $-$0&00003 \\ 
2062&8289(7)$^b$     & $-$1982 &  $-$0&00008 \\ 
2295&8782(4)$^d$     & $-$1342 &  $-$0&00003 \\ 
2298&7914(1)$^d$     & $-$1334 &     0&00006 \\ 
2411&3156(1)$^{eg}$  & $-$1025 &     0&00517 \\ 
2412&4078(22)$^{eg}$ & $-$1022 &     0&00495 \\ 
2413&4958(4)$^e$     & $-$1019 &     0&00053 \\ 
2424&7839(6)$^c$     &  $-$988 &     0&00031 \\ 
2456&8279(4)$^d$     &  $-$900 &     0&00004 \\ 
2673&85488(14)$^d$   &  $-$304 &  $-$0&00009 \\ 
2705&5359(3)$^e$     &  $-$217 &     0&00079 \\ 
2727&3837(4)$^e$     &  $-$157 &     0&00023 \\ 
2784&553370(25)$^f$  &     0   &     0&00000 \\ 
2809&6788(1)$^d$     &     69  &  $-$0&00019 \\ 
3830&7256(4)$^b$     &   2873  &  $-$0&00040 \\
\hline
  \end{tabular}
\end{center} 
\end{table}

\begin{figure}[!t]
  \centering
  \includegraphics[angle=270,width=0.45\textwidth]{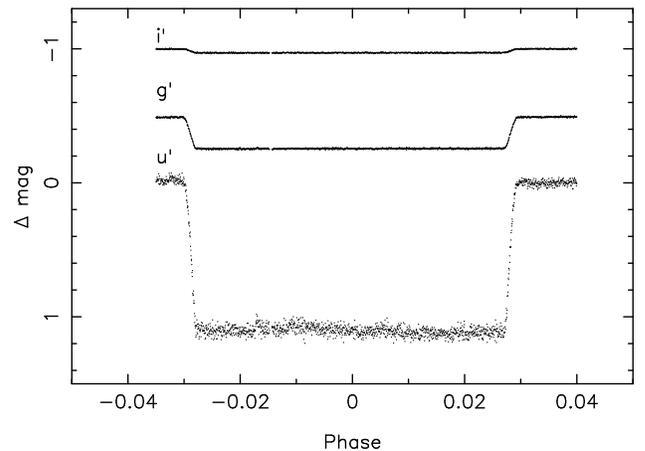}
  \caption{Primary eclipse observed with ULTRACAM.}
  \label{fig:ultracam_eclipse}
\end{figure}

We observed a primary eclipse simultaneously in $u'$, $g'$ and $i'$
with ULTRACAM. These data show the largest eclipse depth of
$1.11\pm0.04$ magnitudes in $u'$. The eclipse depths in $g'$ and $i'$
are $0.235\pm0.004$ and $0.028\pm0.004$ magnitudes respectively, where
we have fitted straight lines to the out of eclipse points and in
eclipse points to derive these values. The difference is taken as the
eclipse depth. The uncertainties are derived for $\Delta\chi^2=1$
when we scale the individual errors to obtain that the reduced
$\chi^2=1$.

\subsection{Apparent magnitudes}
\label{sec:result:mag}
\begin{table*}[!t]
\centering
\caption{$u'$, $g'$ and $i'$ magnitudes of \object{DE~CVn} and the two
comparison stars as derived from the ULTRACAM data taken on May 24,
2003. The uncertainties in $u'$, $g'$ and $i'$ are $0.01$, $0.01$ and
$0.02$, respectively. The magnitudes and colours as derived from the
MDM photometry are given as well with the uncertainty on the last
digits in parentheses.}
\label{tab:ultracammag}
\begin{tabular}{l l l l l l l l}
\hline\hline
  & \multicolumn{1}{c}{$u'$} & \multicolumn{1}{c}{$g'$} & \multicolumn{1}{c}{$i'$} & \multicolumn{1}{c}{$V$}& \multicolumn{1}{c}{$U-B$}& \multicolumn{1}{c}{$B-V$}& \multicolumn{1}{c}{$V-I$} \\
  \hline
DE~CVn in eclipse & 16.43 & 13.74 & 11.65 & - & - & - & - \\
DE~CVn out-of-eclipse & 15.31 & 13.50 & 11.62 & 12.908(2)  &   0.070(15) &   1.263(5)  &   2.244(2) \\
comparison 1 & 15.16 & 13.22 & 12.29 & - & - & - & - \\
comparison 2 & 15.77 & 13.61 & 13.34 & 13.418(2)  &   0.155(13) &   0.708(5)  &   0.784(4)\\
\hline
\end{tabular}
\end{table*}

 In Table~\ref{tab:ultracammag} we give the magnitudes of
\object{DE~CVn} in and out of eclipse of the ULTRACAM data as obtained
with the second comparison star as given in
Sect.~\ref{sec:obs:photo}. Also the MDM magnitudes and the magnitudes
of the comparison stars are given.

The MDM photometry taken on January 21, 2001 was obtained near phase
$0.7$, well outside the eclipse. The transformation to standard
magnitudes was derived using observations of four Landolt standard
fields. Except for $U-B$, the transformations derived from the
standard stars all had scatter below $0.03$ mag. For $U-B$ the scatter
is below $0.08$ mag.

We derived the $u'$, $g'$ and $i'$ magnitudes of \object{DE~CVn} in
and out of eclipse in the ULTRACAM data of May 24, 2003. The part just
before the actual eclipse is taken for the out of eclipse
magnitudes. All the ULTRACAM magnitudes were derived by using the two
comparison stars in the field as mentioned in
Sect.~\ref{sec:obs:photo}. \citet{smith} values were used to derive
the $u'$, $g'$ and $i'$ magnitudes. The magnitudes for \object{DE~CVn}
were the same using either comparison star within $\sim$$0.01$~mag in
$g'$ and $i'$ and $\sim$$0.02$~mag in $u'$ ($1\sigma$ uncertainty).

\subsection{The nature of the components}
\label{sec:result:nature}

\begin{figure*}[!t]
  \centering
  \includegraphics[angle=270,width=0.9\textwidth]{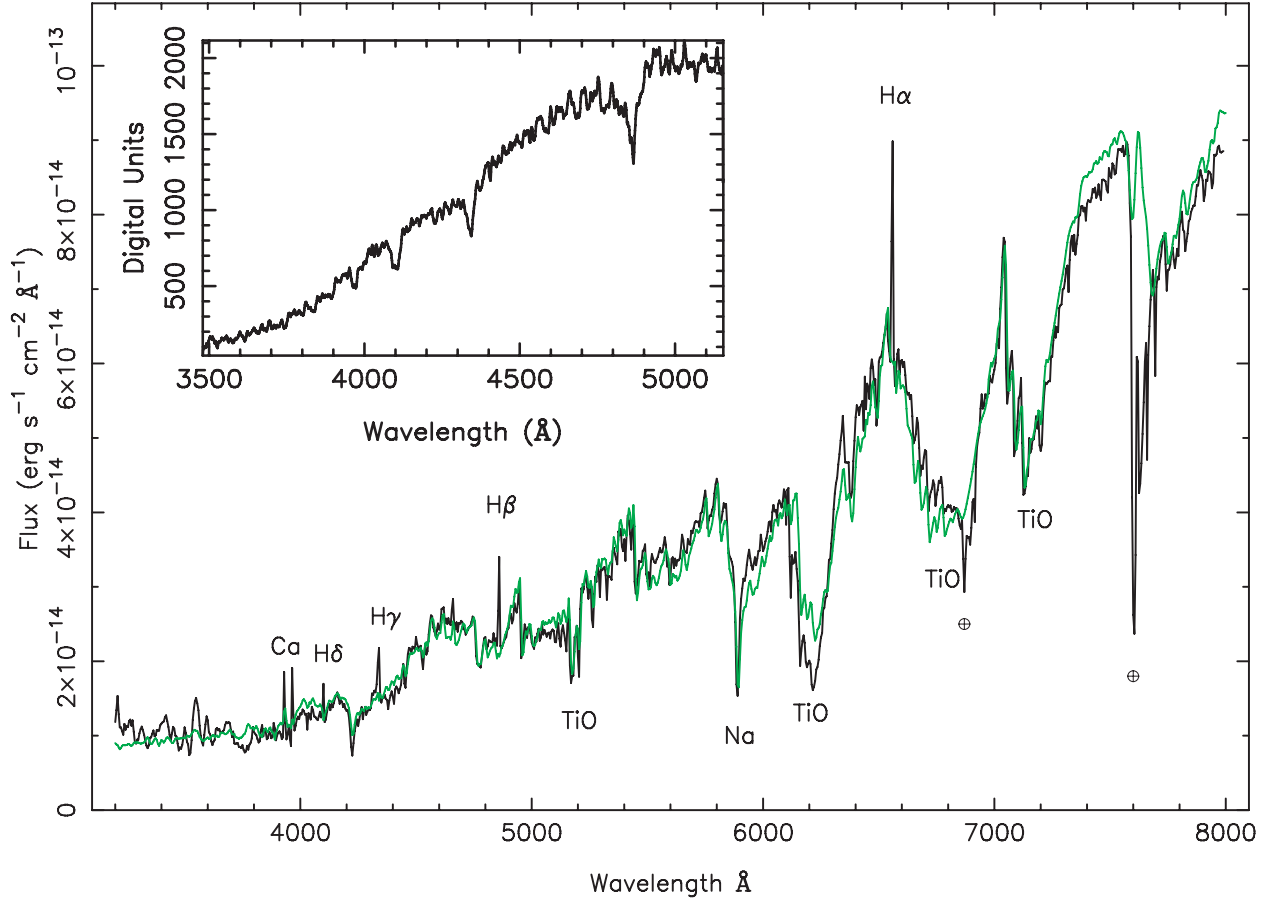}
  \caption{A combined ISIS spectrum of \object{DE~CVn} (black line)
together with a composite template of an M3V star from \citet{pickles}
and a DA white dwarf with a temperature of $8\,000$~K and
$\log{g}=7.5$ (grey/green line). Emission lines of H$\alpha$,
H$\beta$~, H$\gamma$~and H$\delta$~are visible in the spectrum as well
as the \ion{Ca}{II}~H\&K emission lines and the \ion{TiO}{} absorption
bands of the M dwarf. No spectral line signatures of the primary are
visible. The difference between the in and out of eclipse
spectra shows the underlying white dwarf which is plotted in the upper
left corner. }
  \label{fig:spectype}
\end{figure*}

\object{DE~CVn} has not been studied spectroscopically before. An
average low resolution spectrum taken with the ISIS spectrograph on
the WHT on La Palma is shown in Fig.~\ref{fig:spectype}. Clearly
visible are the absorption bands of \ion{TiO}{} indicating an M-type
star. Emission lines of H$\alpha$, H$\beta$, H$\gamma$, H$\delta$~and
\ion{Ca}{II}~H\&K are visible as well.

\object{DE~CVn} is a single-lined spectroscopic eclipsing binary. We
do not see any spectral features of the white dwarf in the
overall spectrum. The six low dispersion DAO spectra referred to in
Table~\ref{tab:speclog} were observed consecutively before, during and
after an eclipse of the white dwarf by the red dwarf. The sum of the
two spectra taken during the eclipse were then subtracted from the sum
of the two spectra taken immediately before the eclipse. The
resultant smoothed spectrum is plotted in Fig.~\ref{fig:spectype}.
Using the spectra taken after the eclipse resulted in a similar
spectrum.  The strong hydrogen absorption lines are typical of a DA
white dwarf and by visually comparing our WD spectrum with the ones in
\citet{wesemael} we come to a spectral type of DA$7\pm0.5$ which
corresponds to a temperature of $7500\pm1000$~K. The lack of residual
\ion{Ca}{II}~K and narrow hydrogen emission lines gives us confidence
that the subtraction was done correctly and the spectra did not need
to be scaled.

To determine the characteristics of the two components we fit the
averaged ISIS spectrum with a composite model consisting of a white
dwarf and a red dwarf. We first corrected the spectra for the radial
velocity variations as a function of phase before averaging the ISIS
spectra.

The comparison spectra that were used to fit the data consist of a
white dwarf model spectrum with a hydrogen atmosphere and temperatures
between $1\,500$ and $17\,000$~K \citep[kindly made available to us by
P. Bergeron:][]{bergeron1,bergeron2}. From the most likely white dwarf
mass and radius as derived from the eclipse fitting in
Sect.~\ref{sec:result:sysparam} we derive a surface gravity for the
white dwarf of $\log{g}\sim$$7.8$. Therefore we have used only white
dwarf template spectra with surface gravity $\log{g}=7.5$. A red dwarf
template (M0V to M6V in integer types) together with the corresponding
absolute visual magnitude was taken from the library of
\citet{pickles}. We first scaled the individual spectra to $10$ pc
before adding them together. We calculated the reduced $\chi^2$ of the
fit to the average ISIS \object{DE~CVn} spectrum for all the different
model composite spectra to determine the nature of the components in
this binary. The composition of the model spectrum with the lowest
reduced $\chi^2$ is taken as the best combination. A more extended
description of the fitting procedure will be given in a future paper.

For the fitting of \object{DE~CVn} we have excluded the wavelength
regions around the emission lines of H$\alpha$, H$\beta$, H$\gamma$,
H$\delta$ and Ca~II~H\&K and the earth atmosphere bands. The best fit
consists of the combination of a red dwarf with spectral type M3V and
a white dwarf with a temperature of $8\,000$~K although the
corresponding formal reduced $\chi^2$ is high ($511$). The second best
fit has a $\Delta\chi^2=50$. When we take all the combinations with a
$\chi^2<1\,000$ the spectral type of the red dwarf stays the same,
while the temperature varies between $7\,000$ and
$9\,000$~K. Therefore we take an uncertainty on the temperature of
$1\,000$~K. If we used $\log{g} = 8.0$ instead of $\log{g}=7.5$ the
spectral type of the secondary stayed the same, while the temperature
changed to $10\,000\pm1\,500$~K. Combined with the temperature
estimate from the eclipse (end of this section) and the difference
spectrum (see above) we decided to use the results from fitting with
the $\log{g}=7.5$ models.

The spectrum of \object{DE~CVn} together with the best fit is shown in
Fig.~\ref{fig:spectype}. Consistent results are found when fitting the
MDM spectra which cover a shorter wavelength range. The discrepancy
between the data and the fit of the spectrum are likely due to flux
calibration errors, different intrinsic properties of the red dwarf
such as metallicity, and the non-removal of telluric features in our
spectra.

Another method to derive the secondary spectral type is by using the
TiO5 index.  \citet{tio5} have given the best-fit linear relation
between the spectral type of a late-type main-sequence star and its
TiO5 band strength:
\begin{equation}
\label{eq:tio5}
S_\mathrm{p} = -10.775\times \mathrm{TiO}5 + 8.2
\end{equation}
where TiO5 is the band strength as defined by
$F_\mathrm{w}/F_\mathrm{cont}$ with $F_\mathrm{w}$ the flux in the
7126--7135~\AA~region and $F_\mathrm{cont}$ as the flux in the
7042--7046~\AA~ region.

Using this definition we have calculated the TiO5 band strength and
the corresponding spectral type for all the MDM spectra covering the
wavelength range given above. The white dwarf contribution to
this part of the spectrum is very small ($\leq5$\%), so this will not
affect the ratio. The corresponding average S$_{\mathrm{p}}$ is
$2.17$, corresponding to spectral type M2.

From the single M3 spectrum of \citet{pickles} we have calculated the
TiO5 band strength and corresponding spectral type as well. This gives
a strength of $0.55$ and a spectral type of $2.27$. By comparing the
value of the M3 spectrum with the intrinsic variation as seen in
Fig.~2 of \citet{tio5} we see that this value is consistent with the
values derived for \object{DE~CVn}. From this we see that the
different methods are consistent with a red dwarf of spectral type M3
as used by \citet{pickles}.

From the apparent magnitudes of \object{DE~CVn} in eclipse and outside
eclipse we can derive the colour of the unseen white dwarf that is
being eclipsed. This gives $(u'-g')_{\mathrm{WD}}= 0.52\pm0.01$ and
$(g'-i')_{\mathrm{WD}}=-0.26\pm0.02$. By comparing this with the
values from the white dwarf models the $u'-g'$ colour indicates a
temperature of 7\,000--9\,000~K, while the $g'-i'$ colour indicates
6\,000--8\,000~K. This is fully consistent with the spectral
modelling.

\subsection{Spectral line variations}
\label{sec:ew}

The spectra show emission lines of hydrogen up to \ion{H}{10} and
\ion{Ca}{II} H\&K emission. We have searched the normalized TLS
echelle spectra for spectral lines showing radial velocity variations,
either in phase with the Balmer and \ion{Ca}{II} H\&K lines or in
anti-phase. All lines identified are listed in Table~\ref{tab:lines}
together with their equivalent widths (EW). No lines were seen to move
in anti-phase with the Balmer lines. The lines that do not have an EW
value are blended with sky lines so that we can not derive an accurate
value for the EW. The EW of the bluest \ion{H}{} lines were measured
in the average ISIS spectrum for which we first removed the phase
shifts in the individual spectra.

\begin{table*}[!t]
\begin{center}
\caption{Identified lines in the TLS echelle spectra and the ISIS
($^a$) spectrum together with their equivalent widths. The lines that
do not have an equivalent width measurement are blended with sky
lines. $\lambda_\mathrm{obs}$ is the wavelength of the line as
measured in the echelle spectra, while $\lambda$ is the wavelength
corresponding to the given element which we identify with this
line. The typical uncertainty on the EW is $\sim$0.1~\AA.}
\label{tab:lines}
\begin{tabular}{l l l l | l l l l}
\hline\hline
$\lambda_\mathrm{obs}$ (\AA) & $\lambda$ (\AA)& Element & EW (\AA)& $\lambda_\mathrm{obs}$ (\AA)& $\lambda$ (\AA)& Element & EW (\AA) \\
\hline
3797.20$^a$ &  3797.91 & \ion{H}{10} & $-$0.9            & 6764.56 & 6764.13 & \ion{Fe}{I} & 0.3   \\
3832.45$^a$ &  3835.397 & \ion{H}{9} & $-$1.8            & 6768.72 & 6768.65 & \ion{Ti}{I} & 0.4  \\
3887.33$^a$ &  3889.05 & \ion{H}{8} & $-$2.1            & 6773.05 & 6772.36 & \ion{Ni}{I} & 0.3   \\ 
3932.47$^a$ &  3933.67 & \ion{Ca}{II}~K & $-$6.2        & 6777.97 & 6777.44 & \ion{Fe}{I} & 0.3\\ 
3968.10$^a$ &  3970.874 & H$\epsilon$+ \ion{Ca}{II}~H & $-$7.5 & 6806.81 & 6806.85 & \ion{Fe}{I} & 0.3 \\
4101.26$^a$ &  4101.735 & H$\delta$ & $-$2.8            & 6815.83 & 6815 & \ion{TiO}{} & 0.5 \\ 
4340.00$^a$ &  4340.465 & H$\gamma$ & $-$3.7            & 6921.45 & 6920.16 & \ion{Fe}{I} & 0.3\\  
4860.63$^a$ &  4861.327 & H$\beta$ & $-$4.1             & 7055.51 & 7054 & \ion{TiO}{} & 1.5 \\
5094.51 &  5093.646 & \ion{Fe}{II} & 0.4            & 7060.87 & 7059.941 & \ion{Ba}{I} & 0.6\\  
5099 &  5098.703 & \ion{Fe}{I} & $<$0.2             & 7088.90  & 7088 & \ion{TiO}{} & 1.0 \\ 
5164 &  5163.940/5164.70 & \ion{Fe}{II}/\ion{Fe}{I} & $<$0.2 & 7126.37  & 7126 & \ion{TiO}{} & 1.3 \\
5167 &  5167.000 & \ion{TiO}{} &       $<$0.2       & 7148.12  & 7148.147 & \ion{Ca}{I} & 0.4\\ 
5210 &  5209.900 & \ion{Fe}{I} &      $<$0.2        &7326.21  & 7326.146 & \ion{Ca}{I} & 0.6\\    
5230 &  5229.857 & \ion{Fe}{I} &      $<$0.2        & 7400.12  & 7400.87/7400.23 & \ion{Fe}{I}/\ion{Cr}{I}& 0.2\\
5240 &  5240.000 & \ion{TiO}{} &     $<$0.2         &7462.33  & 7461 & \ion{Fe}{I} & 0.3\\
6102.78 &  6101.100 & \ion{K}{IV} & 1.3           & 7590  & 7589 & \ion{TiO}{} & blended \\ 
6122.01 &  6122.219 & \ion{Ca}{I} & 1.8           & 7666  & 7666 & \ion{TiO}{} & blended \\ 
6157.08 &  6159 & \ion{TiO}{}     &  0.9               & 7674  & 7671 & \ion{TiO}{} & blended \\
6381.49 &  6384 &  \ion{TiO}{}    &  0.8               &7701  & 7704 & \ion{TiO}{} & blended \\
6388.01 &  6388.410 & \ion{Fe}{I} & 0.8          &7800.24  & 7800.00 & \ion{Si}{I} & 0.3\\
6393.13 &  6391.214/6391.51 & \ion{Mn}{I}/\ion{Ti}{II} & 0.4  & 8047.31  & 8047.6 & \ion{Fe}{I} & 0.3\\    
6421.11 &  6421.355 & \ion{Fe}{I} & 0.3           & 8182.88  & 8183.256 & \ion{Na}{I} & 1.4\\
6439.01 &  6439.073 & \ion{Ca}{I} & 0.9           & 8194.69  & 8194.79/8194.82 & \ion{Na}{I} & 1.4 \\
6449.78 &  6450.854 & \ion{Ba}{I} & 1.0           & 8325  & 8323.428 & \ion{H}{} & blended\\ 
6462.64 &  6462.566 & \ion{Ca}{I} & 1.2           & 8362  & 8361.77  & \ion{He}{I} & blended\\      
6471.56 & 6472.34 & \ion{Sm}{II} &  0.5           & 8379  & 8377.6  & \ion{Ne}{I} & blended\\  
6494.22 & 6493.78 & \ion{Ca}{I}  &  1.0           & 8382  & 8382.23 & \ion{Fe}{I} & blended\\    
6499.39 & 6498.759 & \ion{Ba}{I} &  0.5           & 8387  & 8387.78 & \ion{Fe}{I} & blended\\     
6562.75 & 6562.76 &  H$\alpha$ & $-$6.7             & 8411  & 8412.97 & \ion{Fe}{I} & blended\\
6572.81 & 6572.781 & \ion{Ca}{I} & 0.5            & 8424  & 8424.14/8424.78 & \ion{Fe}{I}/\ion{O}{I}& blended\\  
6593.79 & 6595.326 & \ion{Ba}{I} & 0.5            & 8662.03  & 8661.908 & \ion{Fe}{I} & 1.0\\ 
6626.22 & 6627.558 & \ion{Fe}{I} &  0.3           & 8674.98  & 8674.751 & \ion{Fe}{I} & 0.3\\   
6651.83 & 6651 & \ion{TiO}{} &      0.4           & 8688.44  & 8688.633 & \ion{Fe}{I} & 0.5\\  
6685.09 & 6681 & \ion{TiO}{} &      0.9           & 8806.51  & 8805.21  & \ion{Fe}{I} & 0.4\\      
6719.14 & 6715 & \ion{TiO}{} &   1.3              & 8824.21  & 8820.45  & \ion{O}{I} & 0.6\\   
6760.15 & 6760.61 & \ion{Fe}{I} & 0.4             & \\
\hline
\end{tabular}
\end{center}
\end{table*}

\subsection{Radial velocity curve}
\label{sec:result:radvel}

\begin{figure}[!t]
  \centering
  \includegraphics[angle=270, width=0.45\textwidth]{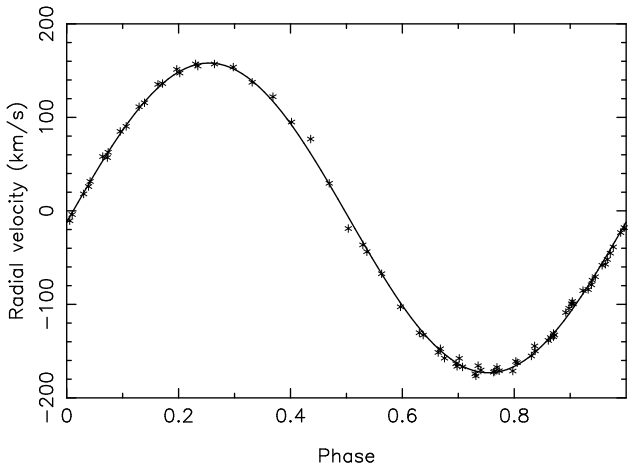}
  \caption{Radial velocity measurements for the H$\alpha$~lines. The
  solid line is the best fit to these velocities, as discussed in
  Sect.~\ref{sec:result:radvel}. The uncertainties on the points are
  smaller than the symbols.}
  \label{fig:radvel}
\end{figure}

\begin{figure}[!t]
\centering
\includegraphics[angle=270,width=0.45\textwidth]{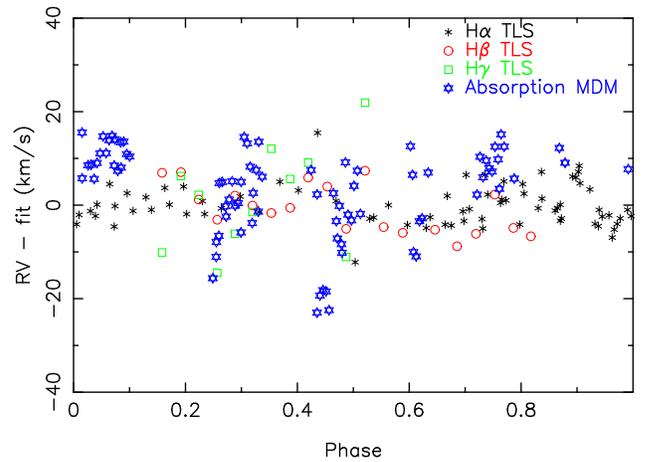}
\caption{Residuals of the H$\alpha$, H$\beta$, H$\gamma$~and the MDM
absorption lines as compared with the best fit sinusoid to the
H$\alpha$~data.}
\label{fig:deltaradvel}
\end{figure}

Radial velocities of the H$\alpha$~lines in the TLS spectra were
determined by fitting a single Gaussian line profile and a first order
polynomial to the emission line and the surrounding continuum. The
radial velocities of the H$\beta$~and~H$\gamma$~lines in the TLS
spectra were also measured in this way. The typical uncertainties on
the radial velocities of the TLS spectra for~H$\alpha$, H$\beta$~and
H$\gamma$~are $\sim$$0.3$ km~s$^{-1}$, $\sim$$0.4$ km~s$^{-1}$ and
$\sim$$0.7$ km~s$^{-1}$, respectively.

The MDM spectra were cross-correlated with an M dwarf spectrum over
the 6000--6500~\AA~range. The uncertainties for these
cross-correlated radial velocities are $\sim$$2.2$ km~s$^{-1}$.

To derive the semi-amplitude of the radial velocity variations of the
secondary and the systemic velocity we use the measurements of the
H$\alpha$~line in the TLS spectra, because these are the most accurate
measurements. There is no spectral line feature of the primary star
visible in the spectra, so therefore we cannot derive the
semi-amplitude of the white dwarf. We have used the function
\begin{equation}
f(\phi)=\gamma+K_2\sin{(2\pi(\phi-\phi_0))}
\end{equation}
where $K_2$ is the semi-amplitude of the secondary star and $\gamma$
the systemic velocity in km~s$^{-1}$ to fit the data. The best fit
gives $\gamma = -7.5\pm3$~km~s$^{-1}$, $K_2 = 166\pm4$~km~s$^{-1}$ and
$\phi_0=-0.004\pm0.004$ for fitting only the TLS
H$\alpha$~variations. The uncertainties are derived from $\Delta\chi^2
= 1$ when the reduced $\chi^2 = 1$.  The phase of the radial velocity
curve is consistent with the time of mid-eclipse.  The fit is shown in
Fig.~\ref{fig:radvel} together with the measured radial velocity
variations of H$\alpha$. The residuals of H$\alpha$, H$\beta$,
H$\gamma$~and the \ion{TiO}{} absorption features are shown
Fig.~\ref{fig:deltaradvel}. When fitting for an eccentric orbit we
find an insignificant eccentricity of $e=0.02\pm0.02$ so assuming a
circular orbit is justified.

No differences in the radial velocity curves of H$\beta$,
H$\gamma$ and absorption with respect to the H$\alpha$ observations are
observed.

\subsection{Irradiation}
\label{sec:irradiation}

\begin{figure}[!t]
  \centering
    \includegraphics[angle=270,width=0.45\textwidth]{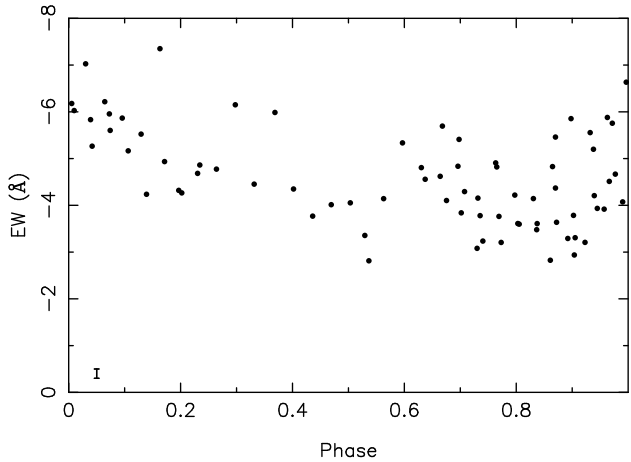}
\caption{The equivalent width of the H$\alpha$~line in the TLS
spectra.  The variations are folded on the orbital period of
\object{DE~CVn}. A typical error bar is shown in the bottom left
corner.}
\label{fig:gaussian}
\end{figure}

In close binary systems we may expect to see irradiation on the red
dwarf due to heating of the close white dwarf. To investigate this
effect we have measured the variations of the EW of the H$\alpha$
line. In Fig.~\ref{fig:gaussian} we show the variations of the EW of
this line folded on the orbital period. A typical error bar is shown
in the bottom left corner of the figure. There is considerable
variation of the H$\alpha$~line, but this does not coincide with the
time scale of the orbital period. During the eclipse the EW is larger
compared to outside eclipse. If the H$\alpha$~emission originates in
the atmosphere of the secondary due to irradiation of the white dwarf,
we would expect to see variations in the strength corresponding to the
time scale of the orbital period.  

In the similar system RR Cae, consisting of a white dwarf of
$7\,000$~K and an M6 or later secondary, a similar effect of a larger
EW during eclipse was seen \citep{rrcae}. This was contributed to the
emission lines being intrinsic to the secondary itself
\citep{rrcae}. In analogy, we therefore conclude that the H$\alpha$
emission in \object{DE~CVn} is also due to activity in the
chromosphere of the M dwarf in contrast to being caused by
irradiation. Furthermore, we can rule out H$\alpha$ emission
emerging from an accretion disk, because of its regular radial
velocity profile and the single peaked lines.

\begin{figure}[!t]
\centering
\includegraphics[angle=270,width=0.45\textwidth]{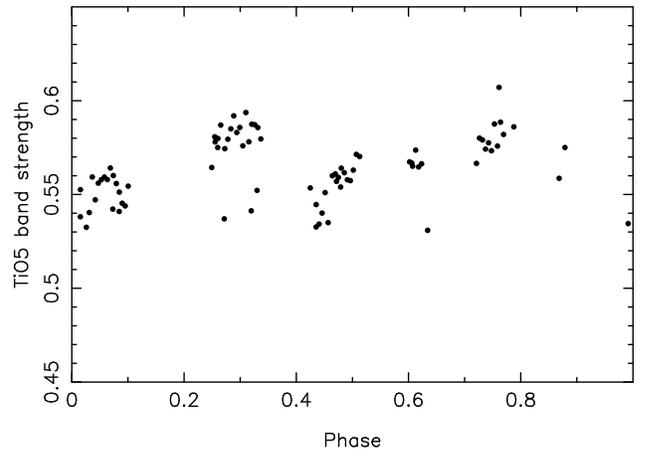}
\caption{TiO5 band strength of the MDM spectra as a function of
orbital phase. The uncertainty on each point is smaller than the
symbol.}
\label{fig:tio5}
\end{figure}

If the secondary star is irradiated by the white dwarf we would also
expect a variation of the TiO5 band strength (which is discussed in
Sect.~\ref{sec:result:nature}) with respect to the orbital period due
to heating in the atmosphere of the secondary. We see variations (see
Fig.~\ref{fig:tio5}), but not with the orbital period. Thus also in
this case, we conclude that the variations are intrinsic to the
secondary.

\begin{figure}[!t]
  \centering
  \includegraphics[angle=270,width=0.45\textwidth]{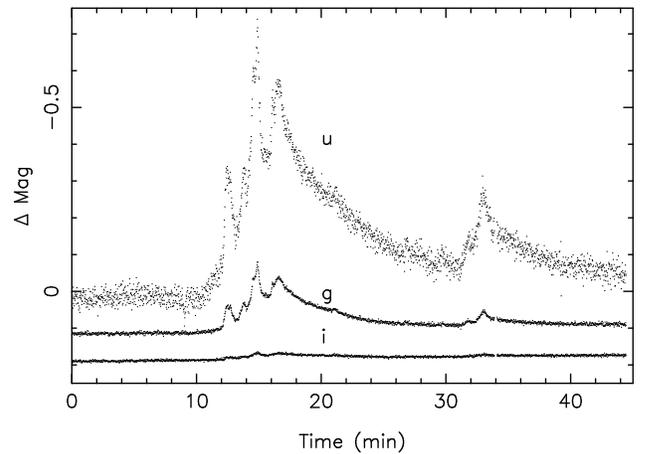}
  \caption{A flare of \object{DE CVn} observed with
  ULTRACAM. }
  \label{fig:ultracam_flare}
\end{figure}

\subsection{Flare}
During one of the ULTRACAM observing runs (25 May 2003) a flare was
observed in all three bands starting at about 23:23 UT. This part of
the light curve is shown in Fig.~\ref{fig:ultracam_flare}. The
observed part of the flare lasted $\sim$$39$ minutes. The flare has a
complex structure with several peaks close together and a
rebrightening during the decay part after the first
peaks. Unfortunately the observation ended before \object{DE~CVn} had
returned to the quiescent (pre-flare) luminosity (most clearly seen in
the $u'$ band). From the observed count rates during quiescence and
the average count rate during the observed part of the flare we
calculated that \object{DE~CVn} increased in brightness on average by
$20$\%, $5$\% and $1$\% of the quiescent flux (so white dwarf + red
dwarf) during this flaring period in $u'$, $g'$ and $i'$ respectively.

\begin{figure}
\centering
\includegraphics[angle=270,width=0.45\textwidth]{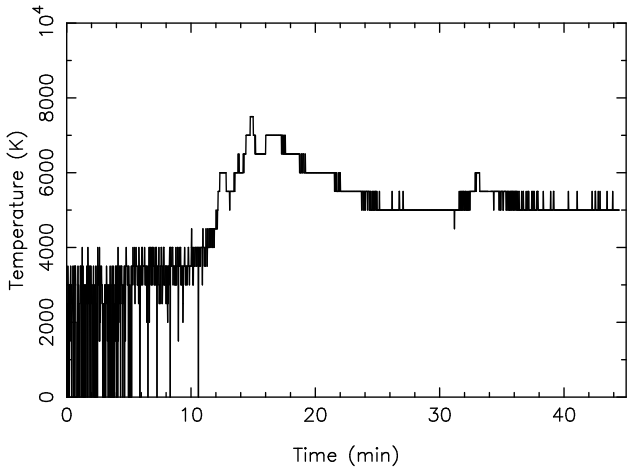}
\caption{Temperature variation of the flaring area during the
flaring period. The flaring area has a radius of $0.038R_\odot$.}
\label{fig:flareTtime}
\end{figure}

To model the flare, we have taken a blackbody spectrum added to the
best fitted template spectrum (see Sect.~\ref{sec:result:nature}). A
blackbody might not be a good approximation, but it is used in other
analyses of flare stars as well \citep{jager1,jager2}. We have assumed
a constant (but not pre-determined) flaring area that is facing us
with only temperature variations over time. We have assumed
temperatures between $0$ and $20\,000$~K in steps of $200$~K and a
constant radius for the flaring area.

For every point in time we have taken the three corresponding observed
magnitudes. We have derived the $u'$, $g'$ and $i'$ magnitudes for our
model by convolving the filter curves with our model spectrum for each
temperature. For each point in time we have calculated the reduced
$\chi^2$ for each temperature and we have taken the temperature with
the lowest reduced $\chi^2$ for this point. After this the average
reduced $\chi^2$ over all the time points was calculated.

We performed this method for several constant flaring areas with radii
between $0.010R_\odot$ and $0.053R_\odot$ with steps of
$0.001R_\odot$. The result with the lowest average reduced $\chi^2$
has a radius $0.038R_\odot$ and the temperature variations are shown in
Fig.~\ref{fig:flareTtime}. The area that flares corresponds to $0.9\%$
of the visible stellar disk of the red dwarf.

\subsection{System parameters}
\label{sec:result:sysparam}
Because spectral features from the white dwarf are not visible we can
only measure the radial velocity curve of the secondary star. The
inclination of the system is constrained by the eclipses in the
photometric light curve.

We have fitted the ULTRACAM $g'$ light curve with a simple model light
curve that results from using two overlapping circles representing a
white dwarf and a red dwarf that is (partly) obscuring the white dwarf
(see Van Ham et al. 2007, submitted). It is assumed that the light
intensity is proportional to the visible part of the white dwarf. We
have used the light curve in counts scaled to range between 1 (out of
eclipse) and 0 (in eclipse). This way we only derive the radii of the
two stars with respect to the orbital separation ($a$,
i.e. $R_{\mathrm{WD}} / a$ and $R_{\mathrm{RD}} / a$).

Limb darkening of the white dwarf could play a role.  When we include
limb darkening in our model the white dwarf radius increased by a few
per cent. This is negligible with respect to the uncertainty in the
inclination so therefore we have neglected the effect of limb
darkening in our method.

From our model we obtain the combinations of the white dwarf and red
dwarf radii with respect to the orbital separation for inclinations
between $75^\circ$ and $90^\circ$ with a step size of $1^\circ$. At
lower inclinations we find that the red dwarf either fills its Roche
lobe or is larger than its Roche lobe. As there is no evidence for
mass transfer in the system, the red dwarf must be smaller than its
Roche lobe. The reduced $\chi^2$ for these fits to the eclipse light
curve are not significantly different, illustrating the fact that
another constraint on the system is required for a unique solution.

The semi-amplitude of the radial velocity variations of the secondary
is derived from the spectra in Sect.~\ref{sec:result:radvel}. The
spectral type of the secondary is determined in
Sect.~\ref{sec:result:nature} by fitting model spectra to the spectrum
of \object{DE~CVn}. These parameters are taken as known input
parameters in our two-circle model.

From the spectral type of the secondary and assuming zero-age
main-sequence masses and temperatures the mass of the red dwarf should
be between $0.3$ and $0.5M_{\odot}$ \citep[see e.g.][]{Mstarmass}. For
every possible inclination we input the mass of the secondary in steps
of $0.01M_{\odot}$ between these values. Furthermore, we use Kepler's
third law and the mass-radius relation for white dwarfs from Eggleton
as quoted by \citet{verbunt}:
\begin{eqnarray}
R_{\mathrm{WD}} = 0.0114\cdot \Bigg( \Big(\frac{M_{\mathrm{WD}}}{M_{\mathrm{CH}}} \Big)^{-\frac{2}{3}} - \Big( \frac{M_{\mathrm{WD}}}{M_{\mathrm{CH}}} \Big)^{\frac{2}{3}} \Bigg)^{\frac{1}{2}} \times \nonumber \\
\Bigg( 1 + 3.5\cdot \Big( \frac{M_{\mathrm{WD}}}{M_{\mathrm{P}}} \Big) ^{-\frac{2}{3}} + \Big( \frac{M_{\mathrm{WD}}}{M_{\mathrm{P}}} \Big) ^{-1} \Bigg) ^{-\frac{2}{3}}
\label{eq:Reggleton}
\end{eqnarray}
where $R_{\mathrm{WD}}$ is in solar radii, $M_{\mathrm{WD}}$ in solar
masses, $M_{\mathrm{CH}}$ = $1.44M_\odot$ and $M_{\mathrm{P}}$ is a
constant whose numerical value is $0.00057M_\odot$.

For every inclination we obtain a value of $R_{\mathrm{WD}}/a$ from
the eclipse fitting. By combining Kepler's third law and
eq.~(\ref{eq:Reggleton}) we derive the mass and radius of the white
dwarf for every inclination and secondary mass. With the radius of the
white dwarf and the result of the eclipse fitting we derive the
orbital separation and the radius of the secondary star.

By assuming a circular orbit, which is justified by the radial
velocity curve, we can constrain the inclination by using the mass
function given in eq.~(\ref{eq:k2}) to derive the semi-amplitude of
the red dwarf for every inclination and secondary mass, and the system
parameters derived from the procedure described above.  With these
parameters and assuming a circular orbit we can use the following
function for the mass function:
 \begin{equation}
 \frac{M_1^3\sin^3{i}}{ \left( M_2 + M_1
 \right) ^2} = \left( 1.0361\times10^{-7}
 \right)~K_2^3~P
\label{eq:k2}
 \end{equation}
 with $M_{1,2}$ in $M_\odot$, $K_{2}$ in km~s$^{-1}$ and $P$ in days.

We take a $3\sigma$ uncertainty on the measured semi-amplitude of the
radial velocity variation of the secondary and compared these with the
semi-amplitudes for each possible solution. We selected only those
solutions that satisfied the radial velocity criterion. This
constrains the inclination of \object{DE~CVn} to $\ge82^\circ$. The
combined radial velocity and eclipse constraints lead to system
parameters where the mass of the white dwarf is between $0.48$ and
$0.65M_\odot$ and radius between $0.0117$ and 0.$0140R_\odot$. The red
dwarf has a mass between $0.30$ and $0.50M_\odot$ and radius between
$0.36$ and $0.51R_\odot$. The orbital separation of the system is
between $2.00$ and $2.25R_\odot$.

By comparing our white dwarf radii with the corresponding masses for
carbon core or oxygen core white dwarfs from \citet{Panei} our minimum
and maximum mass for the white dwarf would be at most $5\%$
larger. Furthermore, all the red dwarf masses that we used in deriving
the system parameters are possible solutions. 

In our ULTRACAM data set, we see hints for small
out-of-eclipse light curve variations and variations between the
different observations, but this data set does not cover the complete
orbital period. In our eclipse fitting procedure we only use the
section between phases $-$0.035 and 0.04, so here the effect of these
variations can be neglected.

\subsection{Space velocity and distance}
The fitting procedure described in Sect.~\ref{sec:result:nature} also
determines a distance to \object{DE~CVn} of $28\pm1$~pc. The
proper motion is known as well (see Sect.~\ref{sec:introduction}) and
the systemic velocity is derived from fitting the radial velocity
curve in Sect.~\ref{sec:result:radvel}. \citet{spacevel} give the
equations to calculate the space velocities ($U$, $V$, $W$) and we use
($U_\odot$, $V_\odot$, $W_\odot$) = ($10.00\pm0.36$, $5.25\pm0.62$,
$7.17\pm0.38$) in km~s$^{-1}$ as the velocity of the Sun with respect
to the local standard of rest as given by \citet{spacesun}. The space
velocities of \object{DE~CVn}, defined as
\begin{equation}
(u, v, w) = (U, V, W) - (U_\odot, V_\odot, W_\odot)
\end{equation}
are ($u$, $v$, $w$)~=~($-15.6\pm0.6$, $-40.7\pm1.5$, $-3.2\pm2.9$)
km~s$^{-1}$. $U$ is defined as positive in the direction of the
galactic center, $V$ is positive in the direction of the galactic
rotation and $W$ is positive in the direction of the galactic North
Pole. The derived space velocities for \object{DE~CVn} are consistent
with being a thin disk object \citep{disk}.

\section{Discussion and conclusions}
\label{sec:conclusions}
\subsection{Binary parameters}
We have analysed various spectroscopic and photometric observations of
the eclipsing binary \object{DE~CVn}. The light curves show total
eclipses of the primary by the secondary star. 

The low resolution average ISIS spectrum is fit with model composite
spectra to determine the spectral type of the M dwarf and the
temperature of the white dwarf. The best fit is derived for an M3V
secondary star and the temperature of the white dwarf (with
$\log{g}=7.5$) of $8\,000\pm1\,000$~K. This is consistent with the
temperature estimates derived from the eclipse depths
(Sect.~\ref{sec:result:mag}) and difference spectrum
(Sect.~\ref{sec:result:nature}), and the spectral type from the
TiO5-index (Sect.~\ref{sec:result:nature}).

From the echelle spectra taken with the TLS we have obtained radial
velocity curves of H$\alpha$, H$\beta$~ and H$\gamma$~and also TiO
absorption bands from the MDM spectra. These curves show a
semi-amplitude for the secondary of $166\pm4$~km~s$^{-1}$.

\begin{figure}
\centering
\includegraphics[angle=270,width=0.45\textwidth]{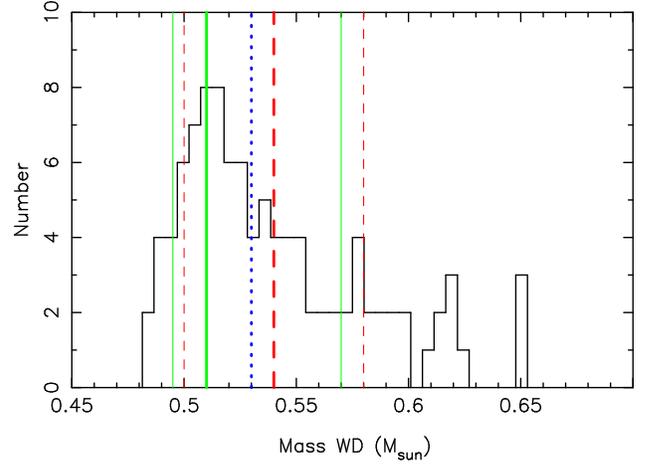}
\caption{Distribution of the possible solutions for the mass of the
white dwarf. The thick solid line (green/grey) is the most likely
value that we have used. The thin solid lines (green/grey) are the
values between which $68\%$ of the possible solutions are located. The
thick dashed line (red/darker grey) is the average value with the
standard deviation in thin dashed lines (red/darker grey). The median
of the distribution is shown as the dotted line (blue/dark grey)}.
\label{fig:masswd}
\end{figure}

Combining the eclipse constraints with the radial velocity amplitude
results in a set of solutions satisfying the constraints as described
in Sect.~\ref{sec:result:sysparam}. Determining the most likely value
with an uncertainty on it is very complex. Due to the interdependency
of the relations used, calculating the uncertainties is not
straightforward. As an example, we show the histogram of the possible
solutions for the white dwarf mass in Fig.~\ref{fig:masswd} which
shows an asymmetrical distribution. The average or the mean does not
coincide with the peak of the distribution. We take the most likely
value as the highest point in the distribution. To determine the
uncertainty on the most likely value we have taken the value for which
$68\%$ of the solutions are enclosed around the most likely value. The
difference between these values is taken as our uncertainty. We have
done this separately for each side of the distribution. For the white
dwarf mass and radius this resulted in
$0.51^{+0.06}_{-0.02}$~$M_\odot$ and
$0.0136^{+0.0008}_{-0.0002}$~$R_\odot$. In the same way, the
distributions of the red dwarf mass and radius result in
$0.41\pm0.06$~$M_\odot$ and $0.37^{+0.06}_{-0.007}$~$R_\odot$
respectively. The orbital separation is
$2.07^{+0.09}_{-0.04}$~$R_\odot$ and the inclination is
$86^{+3~\circ}_{-2}$. These most likely values are consistent with
each other within the uncertainties.

\begin{table*}[!t]
\caption{Parameters of different white dwarf progenitors at the time
the core mass reaches $0.52M_\odot$. M$_0$ is the initial mass of the
main-sequence progenitor. M$_i$ and R$_i$ are the mass and radius of
the star when it has reached the required core mass. This means that
the first five are AGB stars, while the last one is a red
giant. $t_{\mathrm{evol}}$ is the time between the birth of the binary
and the start of the CE phase, $a_i$ and $P_i$ are the initial orbital
separation and initial orbital period, $\alpha_\mathrm{CE}$ is the
fraction of the orbital period used for ejection of the CE, while
$\lambda$ is a numerical value dependent on the structure of the
star. The time from main-sequence binary to a CV is given as
t$_{tot}$.}
\label{tab:progenitor}
\centering
\begin{tabular}{l l r@{.}l l r@{.}l r@{.}l r@{.}l l}
\hline\hline
$M_0$ ($M_\odot$) & $M_i$ ($M_\odot$) & \multicolumn{2}{c}{$R_i$ ($R_\odot$)} & $t_{\mathrm{evol}}$
(yr) & \multicolumn{2}{c}{$a_i$ ($R_\odot$)} & \multicolumn{2}{c}{$P_i$(d)} & \multicolumn{2}{c}{$\alpha_\mathrm{CE}\lambda$} &
$t_{\mathrm{tot}}$ (yr) \\
\hline
1.051 &    0.923 &  231&165 &    6.51$\cdot10^{9}$ &   513&052 &   1166&530 &  0&03 &  2.61$\cdot10^{10}$ \\
1.191 &    1.085 &  224&726 &    4.95$\cdot10^{9}$ &   483&386 &   1007&611 &  0&05 &  2.45$\cdot10^{10}$ \\
1.319 &    1.230 &  218&584 &    3.87$\cdot10^{9}$ &   459&140 &    890&428 &  0&08 &  2.34$\cdot10^{10}$ \\
1.439 &    1.365 &  212&881 &    3.10$\cdot10^{9}$ &   438&689 &    799&406 &  0&11 &  2.27$\cdot10^{10}$ \\
1.555 &    1.494 &  207&045 &    2.54$\cdot10^{9}$ &   419&770 &    722&411 &  0&14 &  2.21$\cdot10^{10}$ \\
3.400 &    3.399 &   40&222 &    2.74$\cdot10^{8}$ &    71&425 &     35&849 &  5&83 &  1.99$\cdot10^{10}$ \\
\hline
\end{tabular}
\end{table*}

\subsection{Progenitor}
From the current small orbital separation we conclude that there was
most probably a CE phase in the past in which the orbit has shrunken
significantly. At the onset of the CE phase, the initially more
massive star has evolved to a star with giant dimensions. We assume
that the core of the giant star at the start of the CE phase had the
same mass as the present day white dwarf. This can be used to
reconstruct the evolution of \object{DE~CVn} to find the possible
progenitors.

To do this we take single main-sequence stars with masses of 1 to
$8M_{\odot}$ in steps of $0.1M_{\odot}$. We evolve these stars with
the method described in \citet{hurley}. When the core mass ($M_c$) of
the stars has reached the mass of the white dwarf that we observe in
\object{DE~CVn} ($0.51M_{\odot}$) the evolution is stopped. Then we
check if the radius of the star at this point ($R_i$) corresponds to
the largest radius during the evolution up to this point. We assume
that the star fills its Roche lobe in the giant phase.

The most likely mass of the white dwarf of \object{DE~CVn} falls in a
mass range where many progenitor stars cannot fill their Roche lobe
with this core mass. The reason is that on the first giant branch
their core grows to $\sim$$0.48M_{\odot}$, when the helium flash
happens. Then the star contracts while the core mass still
increases. When the star expands again to ascend the asymptotic giant
branch (AGB), the core mass has become larger than the most likely
white dwarf mass in \object{DE~CVn}. However, this is very sensitive
to the core mass: for a core mass of $0.51M_{\odot}$ we find only one
progenitor, while for $0.52M_{\odot}$ we find six. We therefore use
$0.52M_{\odot}$, as a compromise between staying close to the most
likely mass yet allowing as many progenitors as are allowed by the
inferred mass range.

The possible progenitors are given in Table~\ref{tab:progenitor}.  The
first three columns in this table are the initial mass of the
main-sequence progenitor of the white dwarf ($M_0$) and the mass and
radius of the star at the time the evolution is stopped ($M_i$ and
$R_i$). The evolution time of the star until this point is given as
$t_{\mathrm{evol}}$. The first five possibilities are stars that reach
the required core mass while on the AGB, while the last one is
peculiar: it reached the required core mass when it had a
non-degenerate core. If the progenitor was such a star, the system
would have come out of the common-envelope phase as a helium-burning
star (most likely observable as a sub\-dwarf B star) with a low-mass
companion. Only after most helium was burned to carbon and oxygen
would the star have turned into a white dwarf.

There are no possible progenitors with initial masses between
1.6--3.4$M_\odot$. For mass $\mathrm{>1.6}M_{\odot}$ the core mass of
the giant on the red giant branch (RGB) has not yet grown to
the required mass. After the RGB phase the star contracts again, while
the core mass is growing. When the star reaches the AGB phase, the
core mass is already larger than the required mass. This results in no
possible progenitor star. At the moment of sufficient core mass, the
star needs to be larger than any time previously in its evolution
otherwise the star would have started Roche lobe overflow earlier in
its evolution when the core mass was not yet massive enough.

For stars $>3.4M_\odot$ the core of the giant at the start of the RGB
phase is already larger than the required core mass, leaving no
possible progenitor for the present day white dwarf. The progenitor
with an initial mass of $3.4M_{\odot}$ is just on the edge of being a
possible progenitor for the present day white dwarf in
\object{DE~CVn}.

\subsection{Common-envelope phase}
We assume that a possible progenitor fills its Roche lobe
at the start of the CE phase. By using the formula for
the Roche lobe of \citet{eggleton}
\begin{equation}
 \frac{R_\mathrm{L}}{a} = \frac{0.49q^\frac{2}{3}}{0.6q^\frac{2}{3} + \ln(1
 + q^\frac{1}{3})}
\label{eq:roche}
\end{equation}
(with $q = M_i/M_2$) we can calculate the orbital separation
($a_i$) (and thus the orbital period $P_i$) at the moment the
evolution was stopped. These values are given in
Table~\ref{tab:progenitor} as well.

From the initial orbital period ($a_i$, Table~\ref{tab:progenitor})
and the present day orbital period, we note that a large orbital
shrinkage has taken place. Together with the rather extreme initial
mass ratios, this implies that \object{DE~CVn} went through a
common-envelope phase during its evolution.

We take the parameters of the giants derived above as the ones at the
start of the common-envelope phase. With the use of \citep{dekool2}
\begin{equation}
\frac{GM_i(M_i-M_c)}{R_i} = \alpha_\mathrm{CE}\lambda \Bigg(
\frac{GM_cM_2}{2a_f} - \frac{GM_iM_2}{2a_i} \Bigg)
\label{eq:alphalambda}
\end{equation}
we calculate $\alpha_\mathrm{CE}\lambda$, where $\lambda$ is a
numerical value dependent on the structure of the star and
$\alpha_\mathrm{CE}$ the fraction of the orbital energy that is used
for ejection of the common envelope \citep{dekool}, for every possible
progenitor. $\alpha_\mathrm{CE}$ is usually
$0\leq\alpha_\mathrm{CE}\leq1$, but for double white dwarfs it is
found to be $\geq1$ \citep{nelemans}. The $\alpha_\mathrm{CE}\lambda$
values for the possible progenitor stars are given in
Table~\ref{tab:progenitor} as well.

The most massive progenitor has $\alpha_\mathrm{CE}\lambda$ of
$5.83$. \citet{dewi} give $\lambda$ values as a function of
the radius of the giant for stars with $M\geq3M_{\odot}$. From their
table we conclude that for our possible progenitor of mass
$3.4M_{\odot}$ $\lambda$ is around 0.6--0.9 indicating
$\alpha_\mathrm{CE}\gtrsim6$ which is very unlikely. Therefore we rule
out such a star as the possible progenitor.

The other possible progenitors have $\alpha_\mathrm{CE}\lambda < 1$
indicating a rather inefficient ejection of the CE (e.g. a large
fraction of the binding energy is not used for ejection of the
CE). These values are in agreement with the values found for other
white dwarf - red dwarf binaries \citep[e.g.][]{nelemans,morales}, and
we conclude that the progenitor star of the present day white dwarf
was a main-sequence star with $M\leq1.6M_\odot$.

\subsection{Time scales}
Now that we can reconstruct the evolution of \object{DE~CVn} we can
determine the corresponding time scales. From the possible progenitors
of the white dwarf we find that the time it took before the CE phase
started is 2.5--6.5$\times10^{9}$ years.

The temperature of the white dwarf is $8\,000\pm1\,000$~K. From the
cooling tracks of \citet{wood} as shown in Fig.~4 of \citet{schreiber}
we can determine a cooling age ($t_\mathrm{cool}$) of the white dwarf
of $\sim$$8\times10^8$ years. The current age of the system is
therefore 3.3--7.3$\times10^{9}$ years.

Due to angular momentum loss, the system will evolve towards a
semi-detached cataclysmic variable (CV) phase. To determine the time
that it will take for \object{DE~CVn} to become a semi-detached binary
we first need to know the orbital period at which mass transfer starts
($P_\mathrm{sd}$). This period follows from the Roche geometry and Kepler's
third law:
\begin{equation}
P_\mathrm{sd} = 2 \pi \Bigg( \frac{R_2^3}{GM_2 \big(1+\frac{M_1}{M_2} \big) \big(\frac{R_\mathrm{L_2}}{a} \big)^3} \Bigg)^{0.5}
\end{equation}
For \object{DE~CVn} we derive an orbital period at the start of mass
transfer of $3$ hours.

The orbital period will shrink due to loss of orbital angular
momentum. For low-mass stars it is often assumed that the dominant
mechanism is (disrupted) magnetic breaking
\citep{verbuntzwaan,king}. In short, disrupted magnetic breaking
occurs in a close binary when the tides force the secondary star to be
co-rotating with the binary, while magnetic fields in the secondary
force the stellar wind to co-rotate with the secondary star. When this
exert a spin-down torque on the secondary star this must extract
angular momentum from the binary orbit. At a mass of
$\sim$$0.3M_\odot$ the secondary becomes fully convective and
therefore loses its dynamo (or at least changes it) so that magnetic
breaking is no longer the dominant source of angular momentum loss.

By assuming that the angular momentum loss is due to disrupted
magnetic braking only until the secondary becomes fully convective at
a secondary mass of at least $0.3M_\odot$, the time it will take
before mass transfer starts is \citep{schreiber}:
\begin{equation}
t_\mathrm{sd}^2 =
\frac{2.63\cdot10^{29} G^{\frac{2}{3}}M_1}{(2\pi)^{\frac{10}{3}}(M_1+M_2)^{\frac{1}{3}}}R_2^{-2}\times(P_\mathrm{orb}^{\frac{10}{3}} - P_\mathrm{sd}^{\frac{10}{3}})
\end{equation}
where $M_1$ and $M_2$ are in $M_{\odot}$ and $R_2$ in
$R_{\odot}$. This corresponds to a time of $1.9\times10^{10}$ years
before \object{DE~CVn} becomes a CV, which is just longer than the
Hubble time.

Systems such as \object{DE~CVn} will not contribute to the current
sample of CVs, unless the loss of angular momentum in the current
detached white dwarf - red dwarf phase is much higher than that given
by magnetic braking alone \citep[see e.g.][ where an angular momentum
loss mechanism $\sim$100 times greater in strength than the currently
accepted value seems to be required to explain the rate of period
decrease in \object{NN~Ser}]{brinkworth}.

\begin{acknowledgements}
We thank Pierre Bergeron for making his cool white dwarf models
available to us.

EvdB, LMR and PJG are supported by NWO-VIDI grant 639.042.201 to
P.J. Groot. GN is supported by NWO-VENI grant 638.041.405 to
G. Nelemans.

JRT thanks the U.S. National Science Foundation for support through
grants AST-9987334 and AST-0307413.  Tim Miller and Maddie Reed took
some of the MDM photometric observations, and Bill Fenton took some of the
MDM spectra.

TRM was supported by a PPARC Senior Fellowship during the course of
this work.

ULTRACAM is supported by PPARC grants PP/D002370/1 and
PPA/G/S/2003/00058.

The William Herschel Telescope is part of the Isaac Newton Group of
Telescopes operated on the island of La Palma by the Inst\'ituto de
Astrof\'isica de Canarias on behalf of the British PPARC and the Dutch
NWO.

We acknowledge the use of the 0.5m telescope of the Cimenhage
Observatory and the 1.8m telescope of the Dominion Astrophysical
Observatory located in Victoria, Canada.

We acknowledge the use of the 1.3m telescope and the 2.4m Hiltner
Telescope of the Michigan-Dartmouth-MIT observatory in Arizona and the
2m Alfred-Jensch-Teleskop at the Th\"uringer Landessternwarte.
\end{acknowledgements}

\bibliographystyle{aa}
\bibliography{de_cvn}
\end{document}